\documentstyle[12pt]{article}
\newcommand{\be}{\begin{equation}}
\newcommand{\ee}{\end{equation}}
\newcommand{\bea}{\begin{eqnarray}}
\newcommand{\eea}{\end{eqnarray}}
\newcommand{\I}{\mbox{i}}
\newcommand{\D}{\mbox{d}}
\newcommand{\E}{\mbox{e}}

\begin{document}
\begin{titlepage}
\begin{flushright}
Freiburg THEP-97/33\\
gr-qc/9802003
\end{flushright}
\vskip 1cm
\begin{center}
{\large\bf  QUANTUM-TO-CLASSICAL TRANSITION
             FOR FLUCTUATIONS IN THE EARLY UNIVERSE}
\vskip 1cm
{\bf Claus Kiefer}
\vskip 0.4cm
 Fakult\"at f\"ur Physik, Universit\"at Freiburg,\\
  Hermann-Herder-Stra\ss e 3, D-79104 Freiburg, Germany.

\vskip 0.7cm
{\bf David Polarski}
\vskip 0.4cm
 Lab. de Math\'ematiques et Physique Th\'eorique, UPRES A 6083 CNRS,\\
  Universit\'e de Tours, Parc de Grandmont, F-37200 Tours, France.\\
\vskip 0.3cm
 D\'epartement d'Astrophysique Relativiste et de Cosmologie,\\
 Observatoire de Paris-Meudon, F-92195 Meudon cedex, France.

\vskip 0.7cm
{\bf Alexei A. Starobinsky}
\vskip 0.4cm 
 Landau Institute for Theoretical Physics, \\
 Kosygina St. 2, Moscow 117940, Russia.

\end{center}
\vskip1cm
\small
\begin{center}
{\bf Abstract}
\end{center}
\begin{quote}
According to the inflationary scenario for the very early Universe, all  
inhomogeneities in the Universe are of genuine  
quantum origin. On the other hand, looking at 
these inhomogeneities and measuring them, clearly no specific 
quantum mechanical properties are observed.
We show how the transition from their 
inherent quantum gravitational nature to classical behaviour comes about -- a 
transition whereby none of the successful
 quantitative predictions of the inflationary 
scenario for the present-day universe is changed.
This is made possible by two properties.
 First, the quantum state for the spacetime metric perturbations 
produced by quantum gravitational effects in the early Universe
becomes very special (highly squeezed) as a result of the expansion of the 
Universe (as long as the wavelength of the perturbations exceeds the Hubble 
radius). Second, decoherence through the environment distinguishes the field
amplitude basis as being the pointer basis.
This renders the perturbations
presently indistinguishable from stochastic classical inhomogeneities.  

\end{quote}
\normalsize

\end{titlepage}

Several years ago, one could not even think about the idea that
a thorough understanding of structure formation in the Universe
would necessarily entail a discussion of fundamental questions
in quantum theory. And yet, this is what currently happens.

How does this come about? Scenarios which include an accelerated expansion
of the Universe at an early stage (an ``inflationary phase")
provide a natural, quantitative, mechanism for structure
formation \cite{BG}. The origin of this structure can be traced back
to unavoidable {\em quantum} fluctuations of some scalar field $\phi$.
These fluctuations then
lead -- together with analogous scalar fluctuations in the metric -- to 
anisotropies in the cosmic background radiation.
In addition, there are relict gravitational waves originating
from tensor fluctuations in the metric.
The COBE-mission and future projects such as the Planck Surveyor satellite 
mission
are able to observe these anisotropies and possibly test
the above scenario \cite{BG,HSS}.

Usually, classical properties for a certain system
emerge by interaction of this system with its natural
environment. This process is called decoherence (see \cite{dec}
for a comprehensive review). Its main characteristics are its
ubiquity and its irreversibility.
 One would expect that this process
is also of utmost importance for the fluctuations
 in the early Universe, and in fact detailed investigations
were performed in this respect (see \cite{hu} and the references therein).

On the other hand, the case was made that as a result of the
{\em special} time evolution of the system -- leading to a
highly squeezed quantum state -- no environment is needed in order to make 
quantitative predictions, since no 
difference to a classical stochastic process can be noticed in
the actual observations.
This was first shown for an initial 
vacuum (Gaussian) state \cite{PS1} and then generalised to
initial number eigenstates \cite{PS2}.

The purpose of our Letter is to provide a conceptual clarification
of this question, which in our opinion should settle this issue.

The simplest example, which nevertheless contains the essential features of 
linear cosmological perturbations, is the case
of a real massless scalar field $\phi$
in a ${\cal K}=0$ Friedmann Universe with accelerated expansion
of the scale factor $a$. This also comprises formally the case of
metric perturbations $h_{\mu\nu}$.
 We use in the following the conformal
time parameter $\eta$ which is defined by $\D t=a\D\eta$,
and denote derivatives with respect to $\eta$ by primes.
It is convenient to consider the rescaled variable $y\equiv a\phi$
(the corresponding momenta thus being related by $p_y=a^{-1}p_{\phi}$).

We consider first the ``system" (the modes of the scalar field)
without invoking any environment. The Hamiltonian can be decomposed
into sums for each wave vector ${\bf k}$,
\be \hat{H}=\int\D^3{\bf k} \left(\hat{H}_{\bf k}^1+
       \hat{H}_{\bf k}^2\right)\ , \ee
where
\be \hat{H}_{\bf k}^1= \frac{1}{2}\left(p_{{\bf k}1}^2 + 
     k^2y_{{\bf k}1}^2 +
      \frac{a'}{a}y_{{\bf k}1}p_{{\bf k}1} 
       \right) \ , \ee
and a similar expression holds for $\hat{H}_{\bf k}^2$, with 
the indices ${\bf k}1$ replaced by ${\bf k}2$. (We note that
$y_{{\bf k}1}=\mbox{Re}y_{\bf k}$, $y_{{\bf k}2}=\mbox{Im}y_{\bf k}$,
where $y_{\bf k}$ is the Fourier transform of the field
$y$, which is a complex variable,
 and similar expressions hold for the momenta.)
Since we are dealing in the following with each mode separately,
we denote the variable just by $y$ (and the momentum by $p$).

The dynamical evolution of the quantum modes is governed by the
time-dependent Schr\"odinger equation,
\be \I\hbar\psi'(y,\eta)=\hat{H}\psi(y,\eta) \ . \ee
What is the initial condition for (3)? The usual assumption is that
at an ``early time" $\eta_0$ (near the onset of inflation),
the modes are in their adiabatic ground state. This may be an immediate
 consequence of quantum cosmological initial conditions \cite{HH,CZ}
(see below), but can also be understood as a natural assumption following, 
e.g., from the simple and elegant principle that the
Universe was in a maximally symmetric state sometime in the past \cite{AS}.

Since the initial adiabatic ground state is a Gaussian state
for the wave function, this Gaussian form is preserved during the
time evolution. However, due to the presence of the $yp$-term
in (2) and the accelerated expansion of the Universe, the state
becomes {\em highly squeezed} \cite{GS}. (Because of the
two contributions in (1), this is a two-mode squeezed state,
although we don't spell this out explicitly in our notation.)
The solution to (3) thus reads
\be 
\psi (y, \eta)=\left(\frac{\Omega_R(\eta)}{\pi} \right)^{1/4}
      \exp\left(-\frac{\Omega(\eta)}{2}y^2\right)\ , \label{psi}
\ee
where $\Omega\equiv \Omega_R+ \I\Omega_I$, and explicit results
can be given for special evolutions $a(\eta)$ \cite{PS1}.

Squeezed states are well known from quantum optics and can be
characterised by the squeeze parameter $r$ and the squeezing angle
$\varphi$.  In the cosmological case, the squeezing
is 
\be \Delta y=C_1a, \quad \Delta p=\frac{\vert C_2\vert}{a} \ . \ee
Here, 
\be
C_1=\frac{H_k}{\sqrt{2k^3}}, \quad C_2=-\I\frac{k^{3/2}}{\sqrt{2}H_k}
        \ , \label{C} \ee
where $H_k$ denotes the Hubble parameter evaluated at the time  
when the perturbation ``crosses'' the Hubble radius 
$H^{-1}=\left({a'}/{a^2}\right)^{-1}$ during the 
inflationary stage. Note that in the original, physical, variables one has 
$\Delta\phi=C_1$ and $\Delta p_{\phi}=\vert C_2\vert$, but that for the 
relevant long wavelengths much bigger than the Hubble radius 
for which $k\eta \ll 1,~C_1\gg\vert C_2\vert {\eta}/{a^2}$. 
The squeezing remains very large after 
these large scales modes cross the Hubble radius for the second time during 
the matter or radiation dominated stage of the recent evolution of the 
Universe.
Thus, in this limit, the squeezing is in the momentum (more precisely, in the 
quantity $\hat{p}-p_{cl}(\eta)$).

For a 
perturbation which will now appear on large cosmological scales, one has
$r\approx 100$. This is more than one order of magnitude
larger than $r$-values obtained in quantum optical experiments
\cite{wine} (recall that $r$ is the logarithm of the amplitude). 

An important property of the state (4) is that in the limit
of large squeezing one has $\Omega_I/\Omega_R\gg 1$.
It follows that it then becomes a WKB state par excellence
\cite{PS1,alb}, although extremely broad in the $y$-direction.

In the Heisenberg picture, the system is characterised by the presence
of a ``growing" and a ``decaying" mode in the solutions
for $\hat{y}(\eta)$ and $\hat{p}(\eta)$ \cite{PS1,alb}.
For large times one gets
\be \hat{y}(\eta) \to f(\eta)\hat{y}(\eta_0), \quad
     \hat{p}(\eta) \to g(\eta)\hat{y}(\eta_0) \ . \ee
Since the information about $\hat{p}(\eta_0)$ is completely lost
in this limit, $\hat{y}$ and $\hat{p}$ approximately
{\em commute} at late times. This, of course, is the expression
of the WKB situation in the Heisenberg picture. For the modes
which presently appear on large cosmological scales, the ratio
of the decaying to the growing mode is $\E^{-2r}\propto 10^{-100}$.
In the limit (7) one also has
\be [\hat{y}(\eta_1),\hat{y}(\eta_2)] \approx 0\ , \ee
i.e., $\hat{y}(\eta)$ approximately commutes at different times.

Clearly, the state (4) is a genuine quantum state and is,
because of its broadness in $y$, very different from ``classical"
states such as coherent states (narrow wave packets).
How, then, comes the conclusion that (4) cannot be distinguished
from a classical stochastic process \cite{PS1}?

The point is that observations of the cosmic microwave background
an\-iso\-tro\-pies are measurements of {\em field amplitudes}.
For the state (4), all corresponding expectation values are
-- in the limit of high squeezing -- indistinguishable from
expectation values with respect to a classical Gaussian
phase space distribution \cite{PS1}. One could, of course,
in principle observe the quantum coherence present in (4) by relying
on other observables such as $\hat{p}-p_{cl}(\eta)$.
Such measurements can, however, not be
performed with any of the satellite missions, and they are
even very difficult to perform for similar states in the
laboratory \cite{alb,ZC}.

The condition (8) is the condition for a so-called
{\em quantum nondemolition measurement} (QND) \cite{dec,BVT}. 
This means that an observable obeying (8) allows repeated 
measurements with great predictability. For a harmonic oscillator
system such as (2), the appropriate QND observables are
the ``quadrature phase operators" which only in the limit of high
squeezing (neglection of the decaying mode) become identical
to the field amplitude.
In the ``Copenhagen interpretation" of quantum theory, one could express
this by saying that if a measurement put the system into an
eigenstate of $\hat{y}(\eta_1)$, all future measurements would give
the corresponding eigenstates of $\hat{y}(\eta_2)$,
$\eta_2 > \eta_1$, corresponding to the classical evolution
of the system.

We note that a nice analogy to the above cosmological quantum state
is provided by a system as simple as the free nonrelativistic quantum particle.
Describing this by an initially narrow Gaussian packet,
it becomes broader during the dynamical evolution.
The interesting point is that for large times,
$t\gg mx_0/p_0$, where $x_0$ and $p_0$ denote the initial
expectation values of position and momentum, this Gaussian state
becomes an exact WKB state, and one has $\Omega_I/\Omega_R
\gg 1$. This corresponds to the limit of high squeezing in the
cosmological state (4). Like there, in the limit of large times
one has $[\hat{x},\hat{p}]\approx 0$, but this time it is
the information about the initial position, and not the
initial momentum, that is neglected, since
\[ \hat{x}(t) \to \frac{\hat{p}_0t}{m}, \quad \hat{p}(t)
   =\hat{p}_0\ . \]
The momentum is always a QND variable for the free particle,
but for large times also position becomes such a variable,
since $[\hat{x}(t_2),\hat{x}(t_1)]\approx 0$, like (8) above.
Note that for a different but closely related experiment when we fix some
large value of $x$ and measure the arrival time of a particle which
passed $x=x_0$ at $t=0$, the discussed quasiclassical behaviour of $\hat{x}$  
simply means that the classical arrival time $t=(x-x_0)m/p_0$ is much bigger
than its quantum indeterminacy (recently discussed in \cite{AOPRZ})
$\Delta t \sim \Delta xm/p_0 \sim \hbar/E$, where $E$ is the particle energy.

Up to now we considered our system (the field modes) to be perfectly
isolated. How realistic is such a situation?
{}From other applications of quantum theory one knows that the effect of an
environment can be very large, as far as decoherence is concerned,
in situations where the dynamical influence is completely
negligible \cite{dec}. A well-known example is a dust grain in
intergalactic space which assumes classical behaviour already due to the
weak coupling to the microwave background radiation \cite{dec,JZ}.
In the cosmological context, global gravitational degrees of freedom
become extremely classical due to their universal coupling
to all other fields \cite{dec,Ki}. It is for this reason that the above 
framework of quantum theory on fixed backgrounds is consistent.

A quantitative measure for the entanglement of a system with its
environment is the ``rate of de-separation" \cite{dec}.
This is defined as follows. Assume that at an initial instant ($t=0$)
the total state is a product state,
$\vert\Psi_0\rangle= \vert\psi_0\rangle\vert{\cal E}_0\rangle$,
where states $\{\vert\psi_i\rangle\}$ refer to the system
-- the field amplitudes in the above examples -- 
and states $\{\vert{\cal E}_i\rangle\}$ refer to its environment
(other degrees of freedom to which the system couples).
The rate of de-separation, $A$, is defined as the coefficient
in $w_0(t)=1-At^2$, where $w_0(t)$ is the probability to find
any unentangled state. Therefore,
\be A= \sum_{j\neq0,l\neq0} \vert\langle\psi_j{\cal E}_l
    \vert\hat{H}\vert\psi_0{\cal E}_0\rangle\vert^2\ , \ee
where $\hat{H}$ denotes here the full Hamiltonian. If $A\neq0$, the total
state is no longer a product state, and entanglement occurs.

Large entanglement with unobservable (``irrelevant") degrees of
freedom produces loss of coherence in the system. In this process,
a {\em distinguished} basis emerges for the system (the ``pointer
basis") with respect to which the system exhibits classical
behaviour. The corresponding pointer observable, $\hat{\Lambda}$,
must obey
\be [\hat{\Lambda},\hat{H}_{int}]=0 \ , \ee
where $\hat{H}_{int}$ is the interaction Hamiltonian between system
and environment. If in addition
\be [\hat{\Lambda}(t_1),\hat{\Lambda}(t_2)]=0\ , \ee
the pointer basis is robust in time and defines a {\em bona fide}
classical trajectory. As we have seen above, the second condition (11)
is fulfilled for the field amplitudes $\hat{y}$ because it is
a QND variable, see (8).  What about the first condition (10)?
The answer depends of course on the interaction Hamiltonian
whose precise form depends on the model under consideration.
However, typical and most realistic interactions for the spacetime metric 
perturbations $h_{\mu\nu}$ are due to the non-linearity of the gravitational 
field. They are mainly proportional either to $h^2$ for $k\eta \ll 1$, or 
to $\left(\delta \epsilon / \epsilon\right)^2$ for $k\eta\gg 1$, where 
$\delta \epsilon / \epsilon\propto h(k\eta)^2$ is the energy 
density perturbation. In particular, the latter is a usual non-linearity 
arising in Newtonian gravity. These kinds of interactions do not depend on 
the field momenta. 
Interactions depending on the momenta also exist, e.g. scattering of 
gravitons $g+g\to g+g$, but their strength is exceedingly small even during 
the inflationary era, not to mention nowadays. 
Therefore,
$[\hat{y},\hat{H}_{int}]=0$, which together with (8) leads to $\hat{y}$
defining the classical basis.

To estimate the strength of decoherence, the de-separation
parameter $A$ has to be evaluated for the above cosmological
situation. In the simplest case, the coupling of the field 
amplitudes $\hat{y}$ to other degrees of freedom, denoted
collectively by $\hat{z}$, is of the form $\hat{H}_{int}
\propto g\hat{y}\hat{z}$, with $g$ being a coupling constant.
This will give a lower bound to all other, more realistic,
couplings. If one, for example, starts with a state that is a product
of the squeezed state (4) with a coherent state of its
environment (the $\hat{z}$-variable), one finds
\bea A &=& g^2\left(\cosh 2r+\cos 2\varphi\sinh 2r\right) \nonumber\\
   & \stackrel{r\to\infty,\varphi\to 0}{\longrightarrow} &
     g^2\E^{2r}\gg 1\ . \eea 
For the modes which now appear on large cosmological scales we had
$\E^{2r}\approx10^{100}$, so that decoherence would only be negligible
if one fine-tuned the coupling to $g^2<10^{-100}$, which of course is
totally unrealistic.

We note that for the state (4) the de-separation parameter
is proportional to the width of the Gaussian,
$\Omega_R^{-1/2}=A/2kg^2$. Therefore, one recognizes again 
that large squeezing in $p$ (corresponding to large
broadening in $y$) causes a large entanglement and therefore 
large decoherence for the system. This is not surprising, since it is
known that the most classical states in the sense of decoherence
are the coherent states \cite{dec}, which are very different from the
high-squeezing limit. We also note that the above entanglement
would only be absent for large $r$ if one had $\varphi=\pi/2$,
corresponding to an exact squeezing in the $y$-direction
(the exact opposite case from what is happening).

The difference to the isolated case discussed above is now that even
if one possessed the unrealistic capabilities of observing,
for example, particle numbers instead of field amplitudes,
one would never see any effect of coherences between different
``classical trajectories" $\vert y(\eta)\rangle$ -- they represent
``consistent histories" \cite{dec,GOGH} to an excellent approximation.
The quantum origin of the field modes is only reflected in the initial
conditions.
What is left for observations is the fundamental prediction about the Gaussian
statistics of the initial metric perturbations. Note that this prediction is 
not restricted to vacuum initial conditions but remains valid with excellent 
accuracy for a much  wider class of non-vacuum initial conditions
\cite{PS2}. It is remarkable that this prediction turns out to be in very
good agreement with observations of the CMB temperature angular anisotropy
\cite{Kog}, the spatial distribution of galaxies and their peculiar velocities
\cite{Kof} and the abundance of rich clusters of galaxies \cite{Ost} -- no
definite signs of non-Gaussianity in the initial conditions for the 
perturbations were found.

The presence of the environment and its decohering influence
gives also a perfect justification for the
phenomenological use of the ``Copenhagen interpretation".
However, as in usual quantum mechanics, the ``Copenhagen interpretation'' 
and the notion of a wave function collapse during
 a measurement are not required 
for any actual experiment, so this scenario can be described as well by the
``many-worlds'' interpretation, where decoherence guarantees
the dynamical independence of the various branches.

We want to conclude with some remarks on the relevance of the above
for the arrow of time \cite{Zeh}. The Second Law of thermodynamics
only holds in the inflationary Universe because its evolution started with 
a state statistically very improbable but dynamically allowed -- the 
maximally symmetric state. Quantum cosmology can, in principle, give
a justification for this peculiarity. For example, taking an 
initial state which is unentangled with respect to the various
degrees of freedom, one gets automatically, by solving the
equations of quantum cosmology, a Gaussian ground state which then
becomes highly squeezed during the following evolution
\cite{HH,CZ}. 
At present the Universe is still far from a maximally chaotic, or highly 
entropic, state. The main contribution to the total entropy $S$ of the 
Universe inside the present cosmological horizon produced by space-time 
metric perturbations comes from gravitational waves with 
wavelengths of the order of 1 cm -- the characteristic wavelength of CMB 
photons. This contribution does not exceed the entropy of CMB itself which is 
$S_{\gamma}\approx 4\times 10^{88}h^{-3}
\left ( \frac{T_{\gamma}}{2.73~K}\right )^3
k_B$ (each species of massless or almost massless neutrinos makes a 
contribution $S_{\nu}=\frac{7}{22}S_{\gamma}$ to $S$, the Universe is further 
assumed to be flat with zero cosmological constant),
 where $h=\frac{H_0}{100}$, 
while $H_0$ is the present Hubble constant expressed in km/s/Mpc. So, as was first noted by Penrose \cite{Pen} 
the total entropy of the observable part of the Universe is much smaller than 
the entropy of a black hole with the same total mass 
($\sim 10^{124}k_B$) which is believed to be the maximal possible entropy for a
given mass. This large degree of order still existing in the Universe 
can provide a well-defined arrow of time.

 To summarise, although the observed fluctuations in the cosmic
microwave background radiation are of quantum origin, for us they appear to 
be of a perfectly classical nature, and it is impossible to observe any quantum
coherences between them. The above treatment shows the universal nature
of quantum concepts -- the physics of the early Universe is not more
exotic than the physics of quantum optics. 

\section*{Acknowledgements}
The visit of A.S. to France, when this paper was completed, was supported by 
the Ecole Normale Sup\'{e}rieure (Paris). A.S. also acknowledges financial 
support by the Russian Foundation for Basic Research, grant 96-02-17591, as 
well as by the German Science Foundation (DFG) through grant 436 RUS 113/333/3.
C.K. acknowledges financial support by the University of Tours during his 
visit to Tours, and D.P. acknowledges financial support by the DAAD during
his visit to Freiburg.

\end{document}